\documentclass[9pt,twocolumn,twoside]{opticajnl}
\journal{opticajournal} 
\setboolean{shortarticle}{false}

\usepackage{lineno}

\usepackage{hyperref}
\usepackage{cleveref}
\usepackage{physics}
\usepackage{xcolor}
\usepackage{amsmath,amssymb} 

\newcommand\note[1]{\textcolor{black}{#1}}

\title{Spatio-spectral vector light created by optical activity in rubidium vapor}

\author[1,2]{Richard~Aguiar~Maduro}
\author[3,4]{Riaan~P.~Schmidt}
\author[1]{Mustafa~A.~Al Khafaji}
\author[1]{Craig~J.~A.~Millar}
\author[1]{Sphinx~J.~Svensson}
\author[3,4,5]{Andrey~Surzhykov}
\author[2]{Adam~Selyem}
\author[1*]{Sonja~Franke-Arnold}

\affil[1]{School of Physics and Astronomy, University of Glasgow, Glasgow G12 8QQ, United Kingdom}
\affil[2]{Fraunhofer Centre for Applied Photonics, Fraunhofer UK Research Ltd., Glasgow, United Kingdom}
\affil[3]{Physikalisch-Technische Bundesanstalt, Bundesallee 100, D-38116 Braunschweig, Germany}
\affil[4]{Institut für Mathematische Physik, Technische Universit\"at Braunschweig, Mendelssohnstrasse 3, D-38106 Braunschweig, Germany}
\affil[5]{Laboratory for Emerging Nanometrology Braunschweig, Langer Kamp 6a/b, D-38106 Braunschweig, Germany}
\affil[*]{sonja.franke-arnold@glasgow.ac.uk}

\begin{abstract}
We show that optical activity in atomic media can generate multidimensional correlations in frequency, polarization and spatial degrees of freedom, allowing us to observe frequency shifts as image rotations.
Specifically, we demonstrate a pump-probe scheme, where optical pumping with circularly polarized light generates macroscopic magnetization in an atomic vapor, associated with frequency-dependent circular dichroism and birefringence. A vortex probe beam, characterized by spatially varying polarization, maps this optical activity onto the spatial structure of the transmitted light. Measuring the intensity profile in a suitable polarization component then allows us to perform spatially resolved polarization spectroscopy.
On resonance we observe a rotation with a slope in the order of $98$ mrad per MHz.  
These findings may find applications in image-based laser-locking, high-precision spectroscopy, magnetometry, and the generation of hybrid entanglement.

\end{abstract}

\setboolean{displaycopyright}{false} 

\begin{document}
\maketitle

\section{\label{sec:Introduction}Introduction}


Atomic media exhibit polarization, and frequency-dependent optical responses determined by the selection rules of atomic transitions. These responses underlie a wide range of techniques and applications, including atomic spectroscopy, quantum memories, and atomic sensors, which have so far been studied predominantly with homogeneously polarized light. In particular, polarization spectroscopy exploits the differential interaction of atoms with orthogonal polarization components of a probe beam and has become a widely used tool for laser-frequency stabilization and sub-Doppler spectroscopy.

Vector light beams, by contrast, possess spatially varying polarization and are therefore characterized by intrinsic correlations between polarization and spatial modes. When such beams interact with an anisotropic atomic medium, they provide access to optical responses that cannot be captured with uniformly polarized probes. The interaction of vector light with atomic media has been explored in various contexts to study spatial anisotropy \cite{sternControllingInteractionsSpacevariant2016a, Wang/AVSQS:2020, fatemiCylindricalVectorBeams2011}, vector magnetometry \cite{Castellucci/PRL:2021, Wang2024, Qiu:21, Ramakrishna/PRA:2024}, atomic memories \cite{ye_experimental_2019,zeng_optical_2023,yang_efficient_2025,wang_storage_2025}, cold atomic cloud shaping \cite{bqys-81lw}, investigations of structured light \cite{wangDirectlyExtractingAuthentic2019,wangMeasuringOpticalConcurrence2024}, and the detection of longitudinally polarized light \cite{svensson_visualizing_2025} which opens up opportunities to create multi-parameter correlations between polarization, frequency and spatial modes. These are akin to spatio-temporal vector beams that have recently been generated with birefringent crystals  \cite{Kopf2024}, featuring three nonseparable degrees of freedom (DoF) of space, wavelength and polarization. Our system differs in two fundamental aspects from experiments with nonlinear crystals: The frequency response of atomic transitions is much narrower, hence the description in terms of frequency rather than wavelength, and the atomic response is highly sensitive to external magnetic fields. This system fits in the road map to create many degrees of freedom \cite{He2022}, combining concepts from structured vector light \cite{Rubinsztein-Dunlop/JO:2017} with spatio-temporal optical vortices \cite{Bekshaev2024}.

Using various light-shaping techniques, such as digital micromirror devices (DMDs), spatial light modulators (SLMs) \cite{forbesCreationDetectionOptical2016a}, and vector vortex retarders \cite{qplate1, qplate2} we can design the polarization structure of our probe beams to study the symmetries associated with absorption and dispersion within atomic media. The use of vortex vector beams (VVBs) provides the possibility to investigate the response of a medium to a continuous range of polarizations \cite{wangOpticallySpatialInformation2018,shiMagneticfieldinducedRotationLight2015}. 

\begin{figure*}[t]
    \centering
    \includegraphics[width=.9\linewidth]{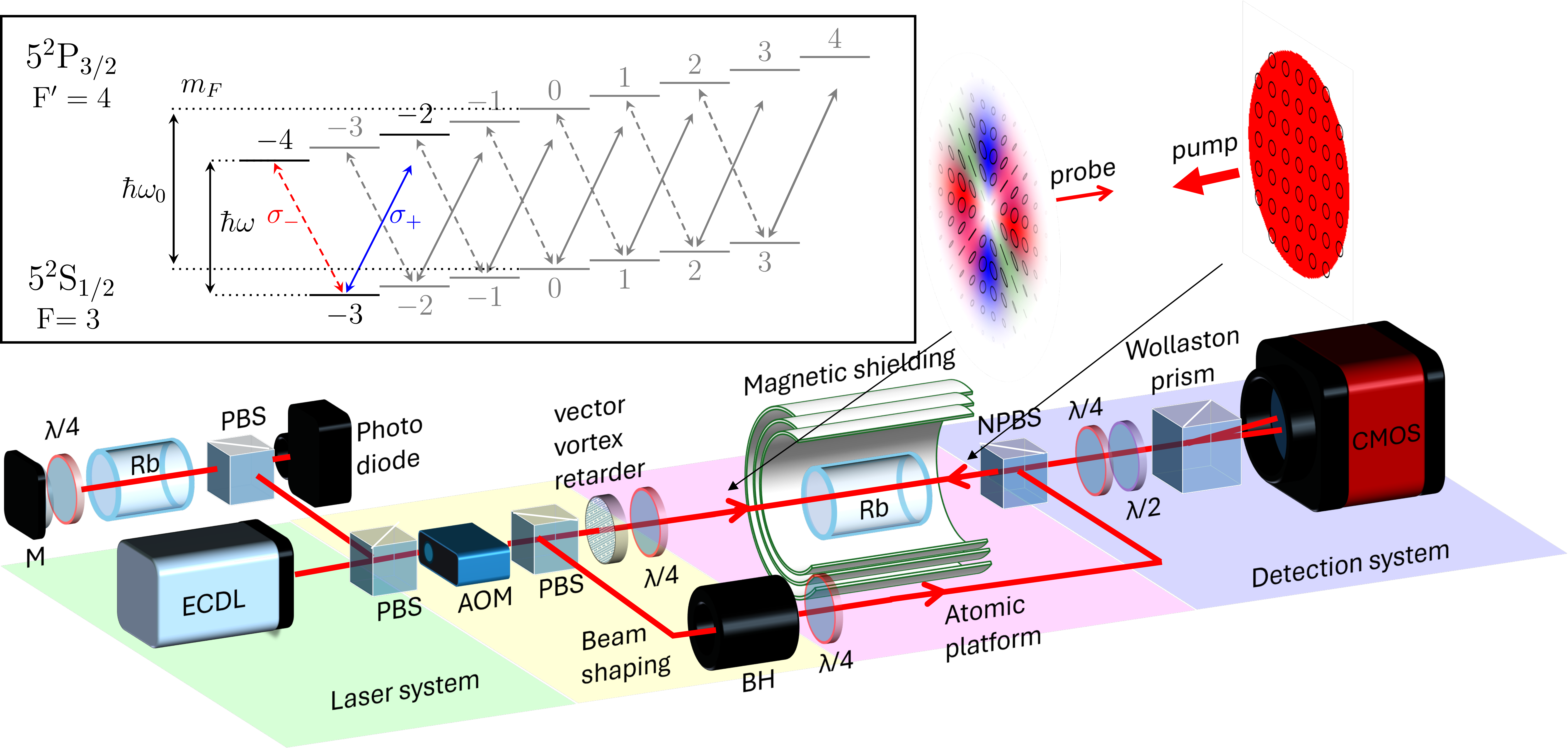}
    \caption{Schematic layout of the spatial polarization spectroscopy setup. A homogeneously polarized pump beam with top-hat intensity profile prepares ${^{85}}$Rb atoms in the $F=3$, $m_F=-3$ ground state, which is probed by a HVB. The polarization structures of the probe (\(|\ell|=1\)) and pump beam are indicated using the colormap explained in \autoref{fig:poincare}. Details of the setup are described in the main text. Inset: Energy level scheme for the 
    D2 line of ${^{85}}$Rb atoms, with transitions addressed by the probe shown in color. BH: beam homogenizer, M: mirror, (N)PBS: (non-)polarizing beam splitter, AOM: acousto-optic modulator.}
    \label{fig:schematic}
\end{figure*} 

Many experiments in atomic physics rely on active frequency stabilization with a spectroscopic signal from a vapor cell. Common techniques include saturated absorption spectroscopy, hyperfine pumping spectroscopy \cite{harocheTheorySaturatedAbsorptionLine1972}, dichroic atomic vapor laser locking (DAVLL) \cite{corwinFrequencystabilizedDiodeLaser1998}, frequency modulation and modulation transfer spectroscopy and, most relevant to our work reported here, polarization spectroscopy \cite{wiemanDopplerFreeLaserPolarization1976, doPolarizationSpectroscopyRubidium2008, almuhawishPolarizationSpectroscopyExcited2021}. 
Polarization spectroscopy relies on differential interaction of the atoms with orthogonal polarization components of the probe light. Just like other techniques relying on counterpropagating beams it is suitable to resolve sub-Doppler spectral signals, however with a limited capture range and potentially an offset compared to the resonant line frequency \cite{pearmanPolarizationSpectroscopyClosed2002}. 

In this work we demonstrate the creation of spatio-spectral light through polarisation dependent optical activity. Correlations between optical frequency and spatial rotation provide a means of frequency detection, with applications in image-based laser locking systems and atomic sensing.

\section{\label{sec:Theory}Theoretical Concept}
We begin with a simple analytical description of the pump-probe scheme. Circularly polarized light optically pumps the atomic vapor, generating macroscopic spin orientation and thereby inducing optical activity. This response is subsequently probed and spatially analyzed by a counter-propagating vector (polarization-structured) beam.
While detailed propagation dynamics including the effects of decoherence and  pump-probe interaction could be assessed within the framework of the density matrix theory equations, our simple analysis captures the main features of the process. We illustrate the concept for the specific transitions realized in our experiments, although it applies more broadly to atomic transitions from a ground state with magnetic hyperfine number $F$ to an excited $F'=F+1$ state. 

A weak axial magnetic field lifts the degeneracy of the magnetic sublevels for the $F=3 \to F'=4$ transition of $^{85}$Rb, as shown in the inset of \autoref{fig:schematic}. A homogeneous (right) circularly polarized pump beam with electric field amplitude \(\mathbf{E}_{\text{pump}} = E_0 \hat{\sigma}_- \), and an intensity above saturation creates a population imbalance within the ground state magnetic sublevels, with the majority of atoms in the stretched state with maximal $m_F$\cite{almuhawishPolarizationSpectroscopyExcited2021}. Solving the Bloch equations as outlined in Refs. \cite{Ramakrishna/PRA:2024, schmidtAtomicPhotoexcitationTool2024b}, predicts a population of 91\% in the $F = 3$, $m_F=-3$ ground-state sublevel. \note{For the small detunings used in our Doppler-free configuration under hyperfine pumping this fraction is even higher.}
In the following we only consider these atoms. 

\begin{figure}[t]
    \centering
    \includegraphics[width=\linewidth]{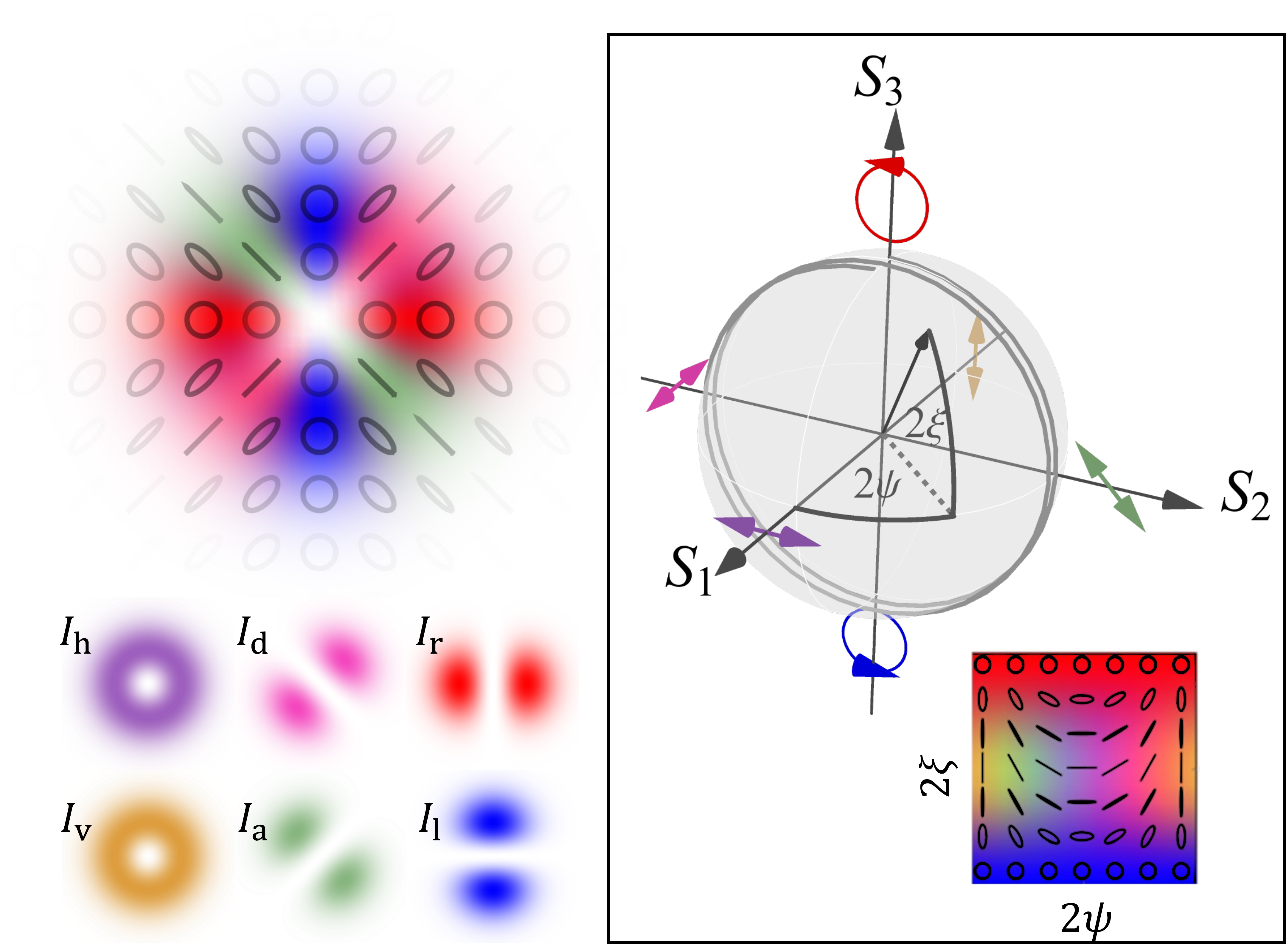}
    \caption{Polarization profile of HVBs of the shape \autoref{eq:probe} used as probe beams, illustrated for the case of $|\ell|=1$. Inset: Polarizations of HVBs are located on a great circle on the Poincar\'e sphere, wrapping around $2|\ell|$ times as a function of the azimuthal angle within the beam profile. In this work we use the indicated color scheme to represent ellipticity $2\xi$ and orientation $2\psi$ of local polarization ellipses within 2D polarization profiles, related to the polar angles of the Poincaré sphere.}
    \label{fig:poincare}
\end{figure}

The \(\hat{\sigma}_\pm\) components of the probe beam can then access V-type transitions from \(m_F\) to \(m_F' =m_F\pm 1\), as indicated in the inset of \autoref{fig:schematic}.  
The macroscopic spin alignment is tested by a counter-propagating probe beam with azimuthally varying polarization structure described by:
\begin{align} 
    {\bf E}_\text{probe}  & =  \frac{E_0}{\sqrt{2}} \left(i\text{LG}^{\ell}_{0}\hat{h} + \text{LG}^{-\ell}_{0}\hat{v}\right) \label{eq:probe} \\
      &=  \frac{E_0}{2} \left[ (\text{LG}^{\ell}_{0} - \text{LG}^{-\ell}_{0})  \hat{\sigma}_+ + (\text{LG}^{\ell}_{0} + \text{LG}^{-\ell}_{0})\hat{\sigma}_-\right], \nonumber
\end{align}
where $\text{LG}^{\ell}_{0}$ are Laguerre Gauss modes with orbital quantum number $\ell$ and radial quantum number $p=0$ \cite{siegman1986lasers}. 
An example of such hybrid vector beams (HVBs) is shown as inset in \autoref{fig:schematic} as well as in \autoref{fig:poincare} for $|\ell|=1$. The latter also includes the representation of HVBs on the Poincar\'e sphere, and their constituent polarization components. We note that the horizontal and vertical polarization component of these HVBs have a donut-shaped intensity profile. 
The second line of \autoref{eq:probe} provides the decomposition of the light into its left and right circular (\(\hat{\sigma}_\pm\)) polarizations which interact with the corresponding $\sigma_\pm$ atomic transitions.
We describe the dielectric response by the atomic susceptibility of a two-level atom,
\begin{equation}
\chi(\Delta_{\rm eff})=\frac{\mathcal{N} \hbar \Omega_R^2}{\varepsilon_0 E_0^2} \frac{i}{\Gamma / 2 - i \Delta_{\rm eff}},
\end{equation}
where the effective detuning $\Delta_{\rm eff} = \Delta + \Omega_{L}$ includes the atomic detuning $\Delta = \omega - \omega_0$ as well as the Larmor shift $\Omega_L$ arising from Zeeman splitting of the ground and excited states. The atomic number density in the interaction region is denoted by $\mathcal{N}$, $\Omega_R$ is the Rabi frequency and $\Gamma$ is the decay rate of the excited state. 

\begin{figure}[t]
    \centering  \includegraphics[width=1\linewidth]{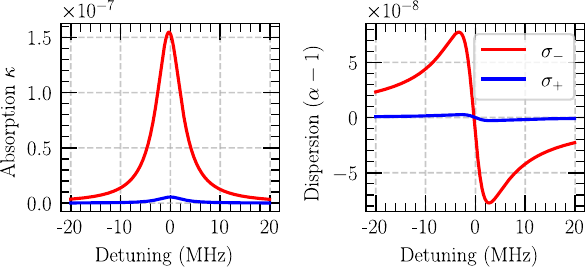}
    \caption{Amplitude absorption and dispersion coefficients for the $\sigma_\pm$ transitions, as given by  \autoref{eq:alpha} and \ref{eq:kappa}  for parameters $\mathcal{N}= 5.38 \times 10^{7} \rm{cm}^{-3}$, $\Omega_{R,+}=0.26$~MHz, $\Omega_{R,-}=1.38$~MHz, $\Omega_{L,+}\approx 0 $, $\Omega_{L,-}=0.294$MHz, $B = 0.21$G and a probe peak intensity $I = 0.173$mW/cm$ ^2$.}
    \label{fig:kramers_kronig}
\end{figure}

The complex refractive index of the atoms, defining both the amplitude absorption coefficient $\kappa$ and the refractive index $n$, is then given by $\tilde n=n +i \kappa=\sqrt{1+\chi}$, which for a dilute medium can be approximated as $\tilde n\approx 1+\chi/2$ \cite{landau, lucarini2005kramers}. For the $\sigma_\pm$ transitions coupling $|F,m_F\rangle \to |F+1,m_F\pm1\rangle$, the Rabi frequencies $\Omega_{R, \pm}$ and Larmor frequencies $\Omega_{L, \pm}$ differ, leading to different absorption and refraction indices for right and left circularly polarized light
\begin{align}
    n_{\pm} &= 1 - \frac{\mathcal{N} \hbar \Omega_{R, \pm}^2}{2\varepsilon_0 E_0^2} \frac{\Delta + \Omega_{L,\pm}}{(\Delta + \Omega_{L,\pm})^2+ (\Gamma/2)^2}, \label{eq:alpha} \\
    \kappa_{\pm}  &= \frac{\mathcal{N} \hbar  \Gamma \Omega_{R, \pm}^2}{4 \varepsilon_0 E_0^2} \frac{1}{(\Delta + \Omega_{L,\pm})^2 + (\Gamma/2)^2}, \label{eq:kappa}
\end{align}
where $\Omega_{L,\pm}=[(m_F\pm 1) g_{F+1}-m_F g_F]\mu_B B/\hbar$ combines the Zeeman shifts of the common ground and the excited states, and $\Omega_{R, \pm}$ are defined by the respective line strengths. \note{As the $\sigma_-$ transition is a stretch-state transition, its Clebsch-Gordon coefficient is much larger.}
We include a plot of \autoref{eq:alpha} and \ref{eq:kappa} for the experimental parameters in \autoref{fig:kramers_kronig}. 

Upon propagation through the atomic vapor, the circular polarization components of the probe beam's electric field then experience different absorption and refraction:
 \begin{multline}
    {\bf E}_\text{probe}  = E(r,z) \left[ i \sin[\ell \varphi] {\mathrm e}^{ikz (n_+ - \kappa_+)} 
    \hat{\sigma}_+ \right.\\
     \left. + \cos(\ell \varphi) {\mathrm e}^{ikz (n_- -  \kappa_-)} \hat{\sigma}_-\right] \label{eq:probe_abs},
\end{multline}
where we have included the radial and $z$ dependence of the LG modes in the amplitude $E(r,z)$.
Differential dispersion, i.e.~circular birefringence, causes a rotation of the polarization ellipse, whereas differential absorption, i.e.~circular dichroism, changes the ellipticity. For polarization structured light, the magnitude of these effects is spatially modulated.
This results in an azimuthal variation of the transmitted beam intensity, driven by circular dichroism:
\begin{multline}
    I \propto \left| {\bf E}_\text{probe} \right|^2 =|E(r,z)|^2
    \left[ \sin^2(\ell \varphi) {\mathrm e}^{-2kz \kappa_+} \right.\\
    + \cos^2(\ell \varphi) {\mathrm e}^{-2kz \kappa_-}\left.  \right]. \label{eq:intensity} 
\end{multline}

Just like for conventional polarization spectroscopy, we project the probe beam onto basis states orthogonal to that of the pump. The intensity $I_{\rm h}$ in the horizontal polarization component is proportional to
\begin{align}
    \left| {\bf E}_\text{probe,h} \right|^2  = &\frac{1}{2}|E(r,z)|^2
     \bigl[ \sin^2(\ell \varphi) {\mathrm e}^{-2kz \kappa_+} + \cos^2(\ell \varphi) {\mathrm e}^{-2kz \kappa_-} \notag \\
    & +  \sin(2 \ell \varphi) \sin(k z(n_- - n_+)) {\mathrm e}^{-k z(\kappa_+ + \kappa_-)} \bigr], \label{eq:intensityXcomponent} 
\end{align}
and $I_{\rm v}=I-I_{\rm h}$, revealing the effect of circular birefringence in the angular position of the azimuthal fringes, and the effect of circular dichroism in the fringe visibility. This effect becomes even more transparent when considering the difference between horizontal and vertical-polarized light, corresponding to the first Stokes parameter:
\begin{multline}
    I_{\rm h} -I_{\rm v} \propto \left| {\bf E}_\text{probe,h} \right|^2 -\left| {\bf E}_\text{probe,v} \right|^2  = \\
    |E(r,z)|^2
     \bigl[  \sin(2 \ell \varphi) \sin(k z(n_- - n_+)) {\mathrm e}^{-k z(\kappa_+ + \kappa_-)} \bigr], \label{eq:S1} 
\end{multline}
For small differences in $n_{\pm}$, the sinusoidal factor $\sin(2 \ell \varphi) \sin(k z(n_{-} - n_{+}))$ in \autoref{eq:intensity} and \autoref{eq:intensityXcomponent} can be approximated by $\sin(2 \ell\varphi+k z (n_{-} - n_{+}))$. The azimuthally modulated intensity pattern thus is rotated by an angle directly proportional to the circular birefringence and scales with  \( |\ell|^{-1}\). As optical activity is frequency dependent, detuning the light from resonance results in a rotation of the observed intensity fringes. 
Equations \ref{eq:intensity}, \ref{eq:intensityXcomponent} and \ref{eq:S1} in conjunction with the frequency and magnetic field dependent absorption and dispersion coefficients in \autoref{eq:alpha} and \ref{eq:kappa} allow us to model the spectroscopic response of our spatial polarization setup.

\section{Experimental setup}
The experimental setup, shown in \autoref{fig:schematic}, allows us to study the interplay between polarization, spatial, and frequency degrees of freedom, enabling spatially resolved polarization spectroscopy of $^{85}$Rb. The rubidium atoms are contained in a $75\,$mm vapor cell at natural isotopic abundance of roughly $72:28$ $^{85}$Rb to $^{87}$Rb. The temperature of the vapor cell is set to $30^\circ$C with a pair of $25$mm Thorlabs glass cell cap heaters and a TC300B temperature controller. \note{We have experimentally established, that the effect prevails for temperatures ranging from $17^\circ\,$C to $45^\circ\,$C, with increased temperatures leading to a slight broadening of the hyperfine absorption lines. We have chosen the operation} temperature to maximize absorption contrast, while staying in the weak probe regime of $I_{\text{probe}} < I_{\text{sat}}$. The vapor cell is surrounded by a three-layer \(\mu\)-metal magnetic shield to restrict the Earth's field lines to the propagation direction of the laser beam, so that the atoms are exposed to an external magnetic field $B_\parallel = (0.21\pm 0.05)\,$G in the Faraday configuration. 

Probe and pump light is generated by an external cavity diode laser (ECDL), which is locked via hyperfine pumping spectroscopy shown in \autoref{fig:schematic} to the $F=3$ to $F'=2,3$ cross-over peak of the D2 line of $^{85}$Rb via a MogLabs DLC102 controller. 
The laser output is then fed to an $80\,$MHz G\&H acousto-optic modulator (AOM) in a double-pass configuration which allows us to change the detuning in increments of $2$ MHz around the atomic resonant frequency, keeping the laser power output constant across the detuning range by modulating the amplitude of the RF signal driving the AOM.

The HVB probe beams described in \autoref{eq:probe} are generated by a combination of polarization optics. A vertically polarized Gaussian beam with $2\,$mm waist and a power of $10\,\mu$W passes through a vortex retarder of $q=1/2$ or $q=1$, for the $|\ell|=1$ and $|\ell|=2$ HVBs respectively, and a quarter-wave plate (QWP) at $45^\circ$ between its fast axis and the horizontal plane. For the $|\ell| =3$ case, a combination of both vortex retarders with a half-wave plate (HWP) between them ensures the OAM has the same handedness to form the correct HVB. \note{While the vortex retarder only affects the phase structure of the probe, the OAM components develop into LG modes (eigenmodes of propagation) in the far field, where the vapor cell is located.} We confirm the polarization structure via Stokes polarimetry \cite{wolfOpticsTermsObservable1954a, stokes}, obtaining the polarization structures for the $|\ell|=1,2$ and $3$ cases shown in the first row of \autoref{fig:rotation}(a).

A counter-propagating right-hand polarized pump beam induces a population imbalance and populates a stretched state, allowing us to access the simple V-level system depicted in \autoref{fig:schematic}. We note that right-hand polarization of the pump beam addresses the $\hat \sigma_-$ transition of the atoms if we assume a quantization axis (coinciding with the direction of $\vec{B}_\parallel$) along the probe propagation direction.
The $1.05\,$mW pump beam is magnified and reshaped by 
a $\pi$ shaper beam homogenizer (BH) to generate a top-hat circular beam with a diameter of $12\,$mm.
The intensity of the probe beam is chosen to be well below the saturation intensity of $I_{\text{sat}} = 1.669\,$mW/cm$^2$ \cite{Steck2008Rubidium8D}, and moreover much weaker than the pump. The pump beam intensity is $I_{\text{pump}} = 1.65\,$mW/cm$^2$ before the cell, and $I_{\text{pump}} = 0.579\,$mW/cm$^2$ after transmission, whereas the incident peak probe beam intensity is $I_{\text{probe}} = 0.173\,$mW/cm$^2$ before and $I_{\text{probe}} = 0.0451\,$mW/cm$^2$ after the vapor cell. This results in a pump-probe ratio of 3.35 to 36.6 immediately before, and after the cell, respectively.

Both pump and counter-propagating probe address the $F=3 \rightarrow F'=4$ hyperfine transition within the D2 line $5^2 S_{1/2} \rightarrow 5^2 P_{3/2}$, with a common detuning $\Delta$ as indicated on the atomic energy level scheme in the inset of \autoref{fig:schematic}. From this ground state, the probe laser can drive V-type $\sigma_\pm$ transitions to $F'=4, m_F'= -2,-4$. 

Spatially resolved polarization spectroscopy is then analyzed by splitting the probe beam into its horizontal and vertical polarization components using a Wollaston prism. This allows us to measure the difference in the frequency dependent circular dichroism shown on the transmitted light. The resulting pattern is captured by a JAI $5000$M CMOS camera with a pixel size of 5$\,\mu\text{m}^2$. 

\section{\label{sec:Results}Results and Discussion}

\begin{figure}[bth]
    \centering
    \includegraphics[width=\linewidth]{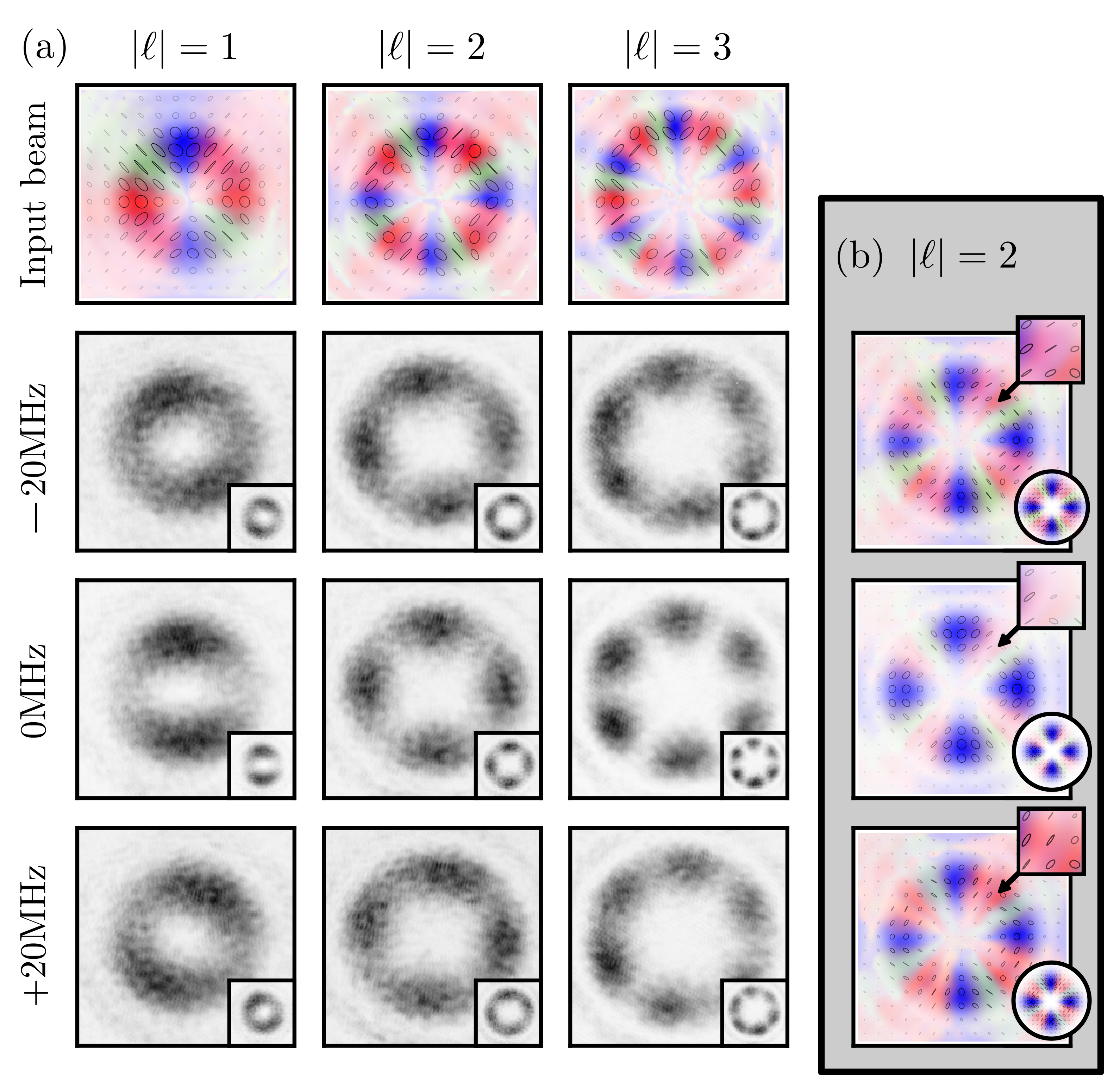}
    \caption{Frequency-dependent image rotation for HVB probe beams with different topological structures. 
    (a) The first row shows the measured polarization profiles of HVBs with $|\ell|=1$, 2 and 3, as defined in \autoref{eq:probe}, with polarizations encoded using the color scheme shown on \autoref{fig:poincare}. The following rows show the transmitted probe intensity after projection onto horizontal (inset: vertical) polarization for laser frequency detunings of $\Delta = -20$, 0 and $+20$ MHz. (b) Experimentally measured frequency dependent Stokes tomography of the transmitted $|\ell|=2$ probe beam for the same detunings, with the theoretical profiles shown in the round insets. 
    The upper insets zoom in to a right circularly polarized area of the input HVB, highlighting the ellipticity shift.}
    \label{fig:rotation}
\end{figure}

Performing spatially resolved polarization spectroscopy with various HVBs of the form given in \autoref{eq:probe} allows us to observe frequency dependent circular dichroism and birefringence. Projecting the input HVB beams into the \(\hat{h},\hat{v}\) basis reveals rings or azimuthally uniform intensity. After transmission through the optically pumped atoms, the intensity shows azimuthal $2|\ell|$ lobes, as shown in \autoref{fig:rotation}(a). As optical activity is frequency-dependent, the position and visibility of these lobes change with detuning.

On resonance, absorption effects dominate. Macroscopic magnetization due to a $\hat{\sigma}_-$ pump light ensures that $\hat{\sigma}_-$ probe light is predominantly absorbed, so that mainly $\hat{\sigma}_+$ components remain in the transmitted beam.
We confirm this by measuring the polarization profile for the example of an $|\ell|=2$ probe beam, shown in \autoref{fig:rotation}(b).

At resonance (middle row), the transmitted beam is predominantly left hand circularly polarized. As $\hat{\sigma}_+$ light comprises equal parts of $\hat{h}$ and $\hat{v}$, projections onto horizontal and vertical polarization show a $2|\ell|$ lobed pattern aligned with the left hand polarized sections of the input beam.   
Away from resonance (top and bottom rows), circular birefringence dominates. This is visible as a Faraday rotation of the locally linear polarization components of the input beam. The effect is highlighted in the square insets of \autoref{fig:rotation}(b), which show that an input polarization oriented along the diagonal is rotated toward the horizontal for negative detuning and toward the vertical for positive detuning.
While circular birefringence does not alter the overall intensity profile of the beam, we may interpret its effect as a rotation of the linear bases of the LG modes in \autoref{eq:probe}, leading to an apparent rotation of the horizontal and vertical intensity components in opposite directions. By looking at the polarisation structure of the total intensity, we see that both the experimental data and simulated beams (shown in the circular insets of \autoref{fig:rotation}(b)) show good agreement in the absorption of $\hat \sigma_-$ light on resonance, and linear rotation off-resonance. 
More generally, the transmitted intensity structure 
is determined by a combination of dispersion and absorption as expressed in \autoref{eq:intensityXcomponent}, with a lobe position determined predominantly by circular birefringence, and lobe visibility by dichroism. See the videos contained in \textbf{Visualization 1-3} for the intensity over the frequency range showed on \autoref{fig:rotation} in which the azimuthal rotation of the lobes can be seen, and \textbf{Visualization 4} for the frequency-dependent polarization structure of the probe beam. 

\begin{figure}
    \centering
    \includegraphics[width=1\linewidth]{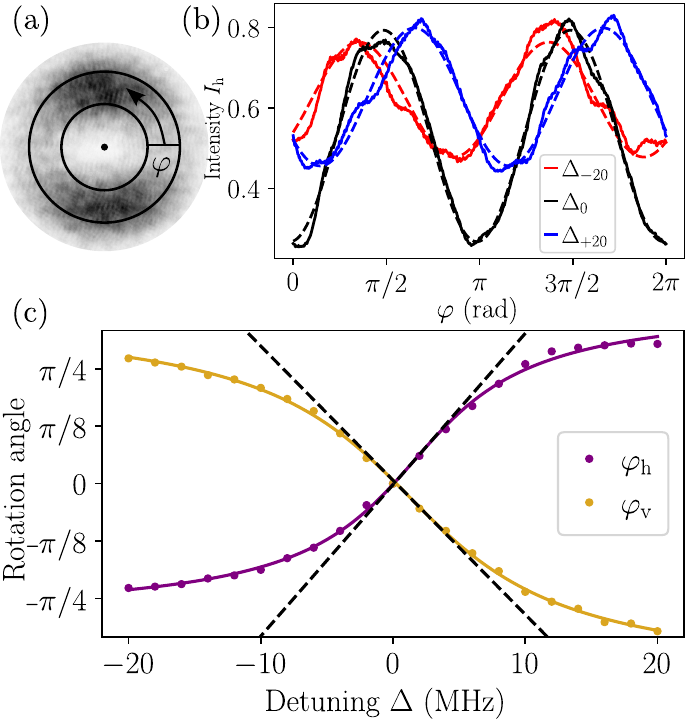}
    \caption{Image rotation as a function of frequency. (a) Conversion of measured intensity profiles into unwrapped intensities. \note{The $\varphi$-dependent intensities are evaluated by averaging over the radius within the indicated ring-shaped area.} (b) $I_\mathrm{h}(\varphi)$ for the example of 0 (solid black line), $-20$ (red) and $+20\,$MHz (blue) \note{detuning}, with sine fits shown as dashed lines.  The rotation angles $\varphi_\mathrm{h}$ and $\varphi_\mathrm{v}$ are then identified from the phase of the sine fit. (c) \note{Measured} image rotation of $I_\mathrm{h}$ (shown in yellow) and $I_\mathrm{v}$ (purple) as a function of detuning, and fitted inverse arctan function (solid lines).
    }
    \label{fig:arctan}
\end{figure}

In order to provide a more complete picture of the frequency dependence of the observed intensity modulation, we have varied the detuning from $\Delta= -20\, {\mathrm{MHz}}$ to $+20 \,{\mathrm{MHz}}$, and taken $21$ snapshots of the transmission pattern in intervals of $2\,$MHz. This frequency range is, up to a small shift of ($3(g_4-g_3)\mu_B B \hbar = 0.294\,$MHz) centered around the \(5^2S_{1/2} (F=3) \rightarrow 5^2 P_{3/2} (F^{\prime} =4)\) transition.  
As the frequency increases, the horizontal projection rotates clockwise, and the vertical projection anti-clockwise. The profiles align at zero detuning. Hence, it is possible to determine the detuning from a single image. For the $|\ell|=1$ hybrid beam the image rotates by $(1.60\pm 0.15)\mathrm{rad}$ over the  observed frequency range. To investigate the rotation, the recorded intensities are Fourier filtered to remove high frequency artifacts due to camera noise. The beam profile is then unwrapped and averaged over the ring-shaped high intensity region indicated in \autoref{fig:arctan}(a). We identify the rotation angles $\varphi_{\rm{h}}$ and $\varphi_{\rm{v}}$ for each detuning by fitting the azimuthal intensity modulation with a sine curve, as shown in \autoref{fig:arctan}(b) for a selection of detunings for the example of $I_{\rm{h}}$. 

Plotting the derived rotation angles as a function of detuning reveals an
inverse tangent function, shown in \autoref{fig:arctan}(c), centered at the resonance frequency of the $F=3, m_F=-3 \to F'=4, m_F'=-3$ transition. The gradient on resonance is $(98 \pm 3)\,\mathrm{mrad/MHz}$, and within the atomic linewidth the rotation angle could be used as a linear feedback parameter for polarization spectroscopy locking.

\begin{figure}[bt]
    \centering
   \includegraphics[width=1\linewidth]{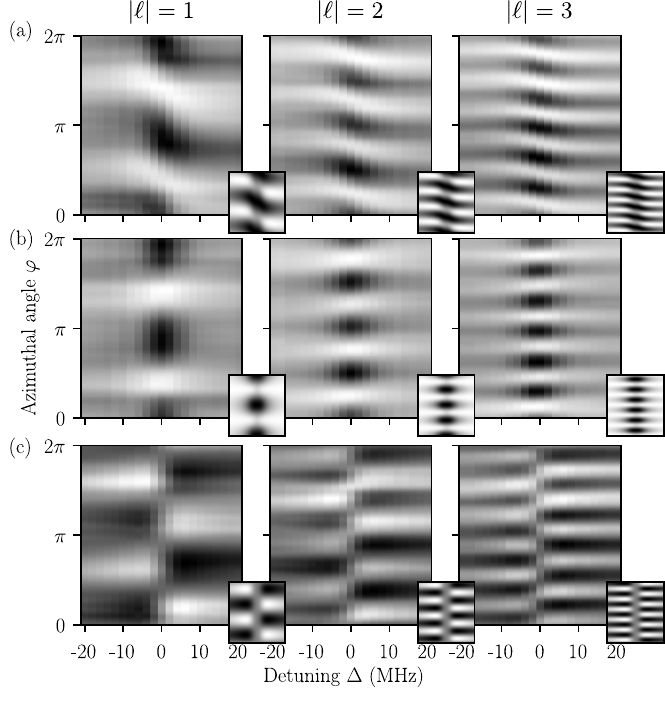}
    \caption{Azimuthal frequency response due to optical activity for HVB probe beans of $|\ell|=1$, 2 and 3, in the left, middle and right column respectively. (a) Image rotation visible as azimuthal shift of $I_{\rm{h}}$ \autoref{eq:intensityXcomponent} as a function of probe beam detuning. Measurements were taken for 21 frequency bands in intervals of $2\,$MHz. The intensity w.r.t. the azimuthal angle $\varphi$ was derived by averaging over the radial parameter as shown in the left inset and is represented by the opacity. 
   (b) Total intensity $I_{\text{tot}} = I_\text h+I_\text v$ \autoref{eq:intensity}. (c) Intensity difference between horizontal and vertical polarization $I_\text h-I_\text v$, as given in \autoref{eq:S1}.
   For each data plot we show the corresponding theoretical profile as inset in the lower right corner.}
    \label{fig:concatenate}
\end{figure}

To visualize the rotation of the transmission profile, we  concatenate the unwrapped and averaged intensities. results scanning over the aforementioned detuning range. 
Measurements and theoretical predictions of $I_\mathrm{total}$ (\autoref{eq:intensity}) and $I_\mathrm{h}-I_\mathrm{v}$ (\autoref{eq:S1}) are reported in \autoref{fig:concatenate}(b) and (c), respectively. The former shows a frequency dependent change of fringe visibility, and the latter an angular shift of $\pi/(2|\ell|)$ of the fringe positions between positive and negative detunings. 

We compare our data to a simulation following the outline given in Section \ref{sec:Theory} for HVBs of the shape \autoref{eq:probe} for our experimental parameters, shown as square insets in \autoref{fig:concatenate}. In all cases we find  excellent qualitative agreement between simulation and experiment.

Taken together, these results illustrate the interplay between circular dichroism and circular birefringence: The lobed intensity pattern close to the atomic resonance originates largely from optical dichroism, as the optically pumped atoms absorb more $\sigma_-$ probe light, visible in $I_{\rm{h}}$ and $I_{\rm{v}}$ (not shown) and the total intensity $I_{\rm{total}}$. At small detunings up to some $\pm \Gamma$, circular birefringence affects the polarization structure of the probe beam, leading to a clockwise (anticlockwise) pattern rotation of $I_{\rm{h}}$ ($I_{\rm{v}})$, visible as pattern shift in $I_{\rm{h}}-I_{\rm{v}}$.

\section{Conclusion and Outlook}

We have visualized correlations between frequency, polarization and spatial degrees of freedom in relation to atomic activity of a Rubidium vapor.  While in our experiments we have concentrated on different HVBs generated as superpositions of linearly polarized LG laser modes with opposite angular momenta, similar effects could be realized for a wide range of structured light beams and magnetic field configurations.

Compared to other media, atoms feature a much narrower frequency response, and optical properties that can be affected by external forces and fields, opening up avenues for enhanced quantum sensing applications.  
We have successfully shown the ability to harness the action of the atomic frequency response for spectroscopy of the narrow magnetic energy sublevels. We illustrate the effects of frequency-dependent dichroism and birefringence in the atom light interaction with atomic media, thanks to the flexibility provided by this multi-parameter system. The symmetry of the system is therefore such that projection into the component Laguerre-Gauss modes of the probe beam results in an apparent spatial rotation caused by the interplay of dispersive and absorptive effects. The measured signal should enable laser locking via image rotation in the vicinity of spectral lines with a small capture range, with the potential to offer off-resonance locking for a small locking range by dynamically applying an external magnetic field. \note{We will explore the generalized effect of arbitrary magnetic fields on spatial image rotation in a future publication.} The flexibility of the additional degree of freedom allows us to design the polarization pattern for specific applications, which can enhance resolution and sensitivity.

\note{Experimental performance relies, of course, in precise control of the desired probe polarization pattern and the projection measurements. In addition, our system operates in the Faraday configuration, and hence we have to minimize fluctuations in magnetic field alignment and/or magnitude. As the signal is derived from the differential interaction between right and left circular polarization components, common mode noise is eliminated, and only unwanted effects of optical activity, e.g. due to imperfect phase plates or strain in cell walls, have to be avoided.}

In summary, we have demonstrated the use of HVBs for rapid, single-shot frequency measurement, as well as the generation of spatio-spectral vector beams in an optically pumped and optically thin atomic medium. We have presented a theoretical model based on the interaction anisotropy between the electrical susceptibility of the $\sigma_\pm$ components of light, allowing us to explain the experimental results intuitively. Hence, our work serves as a demonstration of how multi-parameter correlations can be used in spectroscopic systems.

\begin{backmatter} 
\bmsection{Funding} This work was supported by the QuantERA II Programme with funding received via the EU H2020 research and innovation program under Grant No. 101017733, EPSRC under Grant No. EP/Z000513/1 (V-MAG); R.M.A. received funding via Fraunhofer CAP, award 322761-01; R.P.S. and A.S. acknowledge funding from the Deutsche Forschungsgemeinschaft (DFG, German Research Foundation) under Project-ID 445408588 (SU 658/5-2) and Project-ID 274200144, under SFB 1227 within project B02, and under Germany's Excellence Strategy, EXC-2123 QuantumFrontiers, Project No. 390837967. 

\bmsection{Acknowledgments}
The authors wish to thank Thomas Clark and Jinwen Wang for valuable conversations that helped shaped the early ideas of this project. 

\bmsection{Disclosures}
The authors declare no conflicts of interest.

\bmsection{Data Availability Statement}
The data that support the findings of this study are openly available from the Enlighten Repository at \url{http://dx.doi.org/10.5525/gla.researchdata.2411}.

\bmsection{Author Contributions}
\textbf{Richard~Aguiar~Maduro}: Data curation, Formal analysis, Writing – original draft, Software, Methodology. \textbf{Riaan~P.~Schmidt}: Formal analysis, Writing – review \& editing. \textbf{Mustafa~A.~Al Khafaji}: Investigation, Software, Formal analysis. \textbf{Craig~J.~A.~Millar}: Investigation. \textbf{Sphinx~J.~Svensson}: Investigation, Supervision, Writing - review \& editing. \textbf{Andrey~Surzhykov}: Writing – review \& editing, \textbf{Adam~Selyem}: Writing – review \& editing, Resources, Conceptualization. \textbf{Sonja~Franke-Arnold}: Conceptualization, Supervision, Visualization, Writing – original draft, Funding acquisition, Resources.

\end{backmatter}

\bibliography{main}
\end{document}